\documentclass[iop]{emulateapj} 
\usepackage{times}
\usepackage{natbib}
\usepackage{rotating}

\newcommand{\nii}{{[N\,{\sc ii}]}}

\newcommand{\hii}{H\,{\sc ii}\rm}

\newcommand{\ha}{{H$\alpha$}}
\newcommand{\oh}{12\,+\,log(O/H)}
\newcommand{\ohe}{\mbox{12\,+\,log(O/H)\,=\,}}

\begin{document}
\title{HII Region Metallicity constraints near the site of  the strongly lensed Supernova ``SN Refsdal" at redshift 1.49}

\author{Tiantian~Yuan\altaffilmark{1}, Chiaki~Kobayashi\altaffilmark{1,2}, Lisa J. Kewley\altaffilmark{1}
}
\altaffiltext{1}{Research School of Astronomy and Astrophysics, The Australian National University, Cotter Road, Weston Creek, ACT 2611, Australia}
\altaffiltext{2}{Centre for Astrophysics Research, Science and Technology Research Institute, University of Hertfordshire, Hertfordshire AL10 9AB, UK}

\begin{abstract}
We present the local \hii\ region metallicity near the site of the recently discovered multiply lensed supernova (SN; ``SN Refsdal") at redshift 1.49. 
``SN Refsdal" is located at the outer spiral arm ($\sim$7 kpc) of the  lensed  host galaxy,  which we have previously reported to exhibit a steep negative galactocentric metallicity gradient.  Based on our  updated near-infrared integral field spectroscopic data,  the gas-phase metallicity averaged in an intrinsic radius of $\sim$ 550 pc surrounding  an \hii\ region  $\sim$ 200 pc away from the SN site is   12 + log(O/H)$_{\rm PP04N2}$ $\le$ 8.67.  The  metallicity averaged over nine \hii\ regions at    similar galactocentric distances  ($\sim$5-7 kpc) as ``SN Refsdal"  is constrained to be 12 + log(O/H)$_{\rm PP04N2}$ $\le$ 8.11.        Given the fortuitous discovery of ``SN Refsdal" in an advantageously lensed face-on spiral, this is the first observational constraint  on the local metallicity environment of an SN site at redshift $z>1$.   
\end{abstract}


\section{INTRODUCTION}
Metallicity at the site of the supernova (SN) explosion provides important clues to the physical origins of the SN(e).  Reciprocally,  the  evolution  and feedback of SNe provide crucial recipes  for  the chemical enrichment models of galaxies across cosmic time \citep[e.g.,][]{Scannapieco06,Kobayashi07,Few14}.   

Generally, SNe can be attributed to either the thermonuclear explosion of a white dwarf in a binary system \citep[Type Ia; e.g.,][]{Hillebrandt00,Parrent14} or the  core collapse (CC) of a massive star   \citep[Type II, Ib/c; e.g.,][]{Heger03,Crowther07,Smartt09}.  Theoretically,  at  low metallicities ([Fe/H]  $\lesssim -1.1$), SNe Ia may be suppressed through the single-degenerate channel as the white dwarf wind would be too weak for the mass to reach the Chandrasekhar limit \citep{Kobayashi98,Kobayashi09}.  For CC SNe,  
   the number ratio of SNe Ib/c to SNe II is expected to increase with progenitor metallicity  due to the dependence of mass-loss rate on metallicity \citep[e.g.,][see also \citet{Eldridge13,Yoon10} for the binary scenario]{Meynet03,Prantzos03,Eldridge07,Limongi09}.

However, the strong theoretical dependence on metallicity has not been confirmed  in observations.   At low redshift ($z<1$), 
observations have shown that  the metallicity distributions of the host galaxies of SNe II and SNe Ia are statistically indistinguishable,  whereas the host galaxies of SN Ib/c favor higher metallicity \citep[e.g.,][]{Prieto08,Modjaz11,Stoll13}. The  class of  abnormally luminous SNe II/Ia favor lower-metallicity environments \citep[e.g.,][]{Childress11,Khan11,Kelly12}.    At high redshift ($z>1$),  measurements for SN host metallicities are still rare and deserve special attention  \citep[e.g.,][]{Frederiksen14,Graur14,Rodney14}.

Note that theoretical models of SNe are based on the metallicity of the progenitor star, whereas in observations, direct measurement for the progenitor star metallicity is only possible for SNe in nearby galaxies.     Most studies are based on  assumptions that the global metallicity of the host galaxy is a good proxy for the SNe progenitor metallicity and that gas-phase oxygen abundance is a good proxy for the stellar metallicity.   Additionally,  the majority of  local spiral galaxies show  a radial metallicity gradient and their gradients could  be significantly steeper at high redshift \citep[e.g.,][]{Jones10b,Yuan11}.    It is therefore quantitatively unclear how much systematic uncertainty  in the observed metallicity and SN-type relation is  introduced by the  approximation of the locally and globally averaged metallicity \citep[e.g.,][]{Stanishev12}.

In the era of integral field spectroscopic (IFS) surveys,  spatially resolved properties at the sites of SNe can now be measured for large samples at low redshift, enabling more accurate and unbiased local environment parameters to be probed for the SN progenitors \citep{Galbany14}.   Measuring spatially resolved metallicities at high redshift remains challenging  because of  the limited angular resolution and signal-to-noise (S/N) \citep[S/N;][]{Yuan13b}.  With a boost of a few tens in flux and area through lensing magnification,  gravitationally lensed targets are  the forerunners of this field  \citep[e.g.,][]{Jones13,Wuyts14}.

The recent discovery of a multiply lensed supernova  at $z${=}1.49 \citep[``SN Refsdal";][]{Kelly14} in the cluster field of 
 MACS\,J$1149.5$$+$2223 provides a rare opportunity to study the local metallicity environment of a high-redshift SN for the first time.  
In this Letter, we update the IFS data from \citet{Yuan11} and place local metallicity constraints  near the explosion site of ``SN Refsdal." 
Throughout this paper, we adopt a flat cosmology with
$\Omega_{M}{=}0.3$, $\Omega_{\Lambda}{=}0.7$, and
H$_{0}{=}70$~km~s$^{-1}$~Mpc$^{-1}$. 
We use solar oxygen abundance 12 + log(O/H)$_{\odot}$=8.69 \citep{Asplund09}.

\begin{figure*}[!ht]
\centering
\includegraphics[trim = 6mm 0mm 6mm 0mm, clip, width=16.5cm,angle=0]{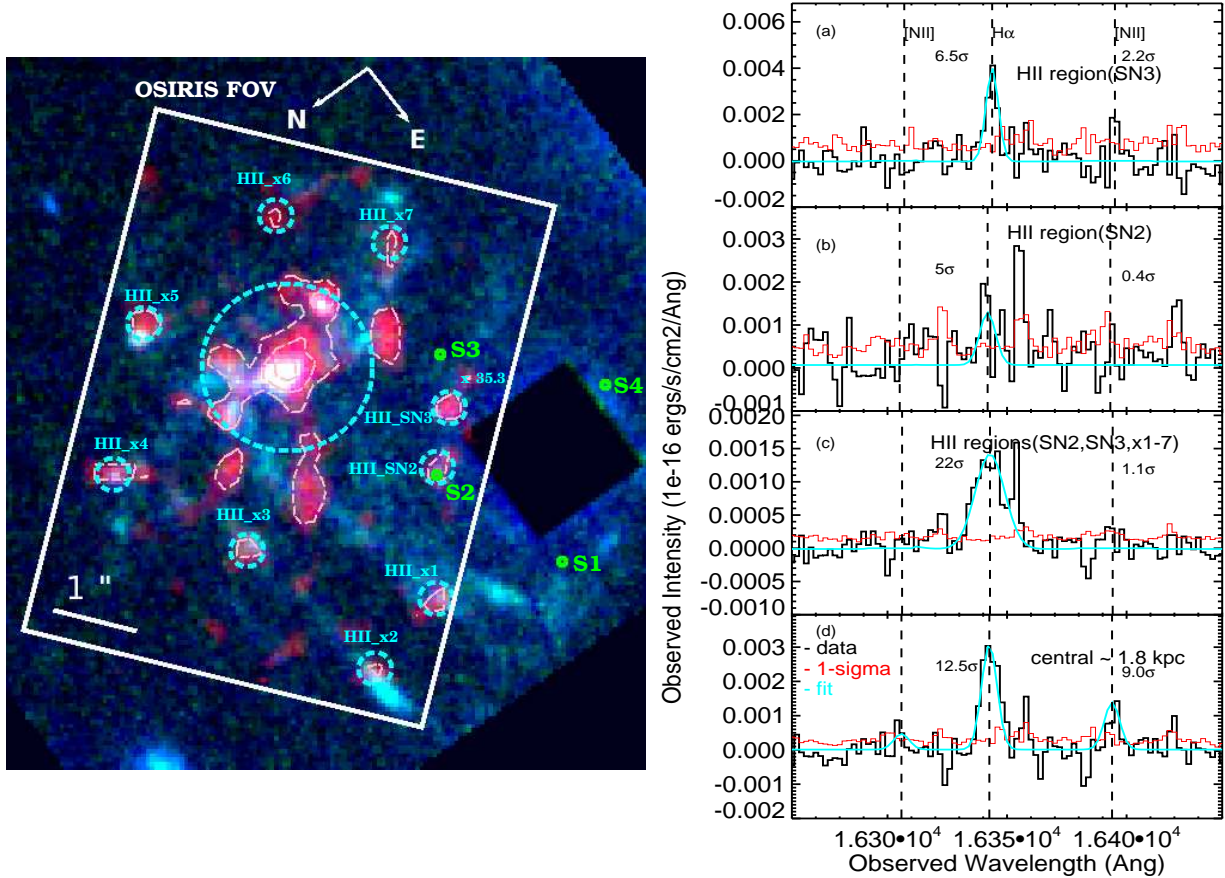}
\caption{Left:   OSIRIS \ha\ (red)  and {\textit HST} F814W (green), F555W (blue) band  color image of the host galaxy of ``SN Refsdal" at  $z=1.49$  behind the lensing cluster MACS\,J1149.5$+$2223.    The white box shows the  4.8 by 6.4 arcsec FOV of the OSIRIS Hn3 filter with the 0.$\!\!^{\prime\prime}$1 per pixel plate scale.    The foreground elliptical galaxy is masked out (black square).  Locations of the ``SN Refsdal" images are highlighted with green circles and are labeled as S1, S2, S3, and S4.   Two  \hii\ regions close to S2 and S3 are labeled as ``\hii\_{\rm SN3}" and ``\hii\_{\rm SN2}."  
Five \ha\ contour levels (0.8, 1.4, 2.2, 2.8, 3.5,  in units of 10$^{-16}$ ergs~s$^{-1}$~cm$^{-2}$~arcsec$^{-2}$) are shown in pink.  
One of the  multiply lensed \hii\ regions defined as associated with the supernova \citep{Sharon14} is labeled as ``35.3." 
Cyan circles  indicate the nine \hii\ regions that are at approximately similar galactocentric distances ($\sim$5-7 kpc) as ``SN Refsdal."   A radius of 0.$\!\!^{\prime\prime}$3 ($\sim$ 550 pc on  the source plane) is used 
to define the \hii\ region size. Right:  averaged spectra for different regions marked on the left.  Raw data are shown in black; the 1$\sigma$ noise spectra are  in red. 
Cyan lines show the best-fit Gaussian profiles for  lines detected above 3$\sigma$ (S/N of the \ha\ and \nii\ lines are marked).  
}
\label{fig:fig1}
\end{figure*}

\section{near-infrared IFS DATA}
Our data were collected on the laser guide star adaptive optics (LGSAO) aided near-infrared integral field spectrograph OSIRIS \citep{Larkin06} at KECK II in 2010 and 2011. 
The data presented in \citet{Yuan11} were taken on 2010  March 3. 
We obtained a total of 4.75 hr net exposure in the Hn3 band on the lensed spiral  galaxy ``Sp1149" ($(\alpha_{2000},\delta_{2000})$\,=\,(11$^{\rm h}$49$^{\rm m}$35.284$^{\rm s}$, +22$^{\circ}$23$^{\prime}$45.86$^{\prime\prime}$)) that is hosting ``SN Refsdal."   
 On  April 22, 2011, we acquired another 2 hr of  net exposure on this source with the same instrumental settings under similar weather conditions.  The data were reduced using the same pipeline and analysis code as described in \citet{Yuan11}.  We coadded the flux-calibrated data of 2010 and 2011 and weighted by the S/N of the total \ha\ flux.

To summarize, the data presented in this paper are based on a  total of 6.75 hr on-source exposure on OSIRIS KECK II.  The spatial resolution in our observation is  $\sim$0\farcs1, corresponding to a source-plane resolution of $\sim$170 pc after correcting for lensing magnification of $\mu_{flux}{=}22$ (using the same lens model  as described in  Yuan et al. 2011). The  spectral coverage is 1.594$-$1.676$\micron$ with a spectral resolution of $\sim$ 3400.  The corresponding rest-frame wavelength range is 6406$-$6731 \AA\
at the redshift of ``Sp1149" ($z{=}1.49$), aiming to capture the  \ha\ and \nii\ emission lines.

Figure~\ref{fig:fig1} shows the two-dimensional (2D) \ha\  map from the OSIRIS datacube for \ha\  line detections of S/N$\ge$ 3$\sigma$.  
 We use the coordinates of the SN sites on the image plane  from \citet{Kelly14} and \cite{Oguri14}.  
The  sites of the lensed supernova  images are marked with green circles (``S1-4"). 
Two SN locations (``S2" and ``S3")  are covered in the OSIRIS field of view (FOV). 
We detect a significant amount ($f$\ha\ $\ga$ 0.8 $\times$ 10$^{-16}$ ergs~s$^{-1}$~cm$^{2}$~arcsec$^{-2}$ or $\ga$ 0.8 M$_{\odot}$ yr$^{-1}$ kpc$^{-1}$) of  \ha\ emission from the \hii\ regions (``\hii\_{\rm SN2}" and ``\hii\_{\rm SN3}") that are close to ``S2" and ``S3." 
Most of the bright \ha\ knots are associated with a strong continuum, though not all continuum peaks have bright \ha\ emissions (Figure~\ref{fig:fig1}).  

Note that ``\hii\_{\rm SN2}" and ``\hii\_{\rm SN3}" are not the multiply lensed \hii\ regions associated with the SN as defined in \citet{Sharon14}. 
We  detect very weak (3$\sigma$) \ha\ emission  from one of the  multiply lensed \hii\ regions (labeled as ``35.3" in Figure~\ref{fig:fig1}) defined in \citet{Sharon14}.
The spectra extracted at the location of ``35.3" using the same aperture size as  ``\hii\_{\rm SN2}" is  similar to ``\hii\_{\rm SN2}" (see Figure~\ref{fig:fig1}, panel (b)).

\begin{figure*}[!ht]
\centering
\includegraphics[trim = 1mm 8mm 5mm 5mm, clip, height=8cm,angle=0]{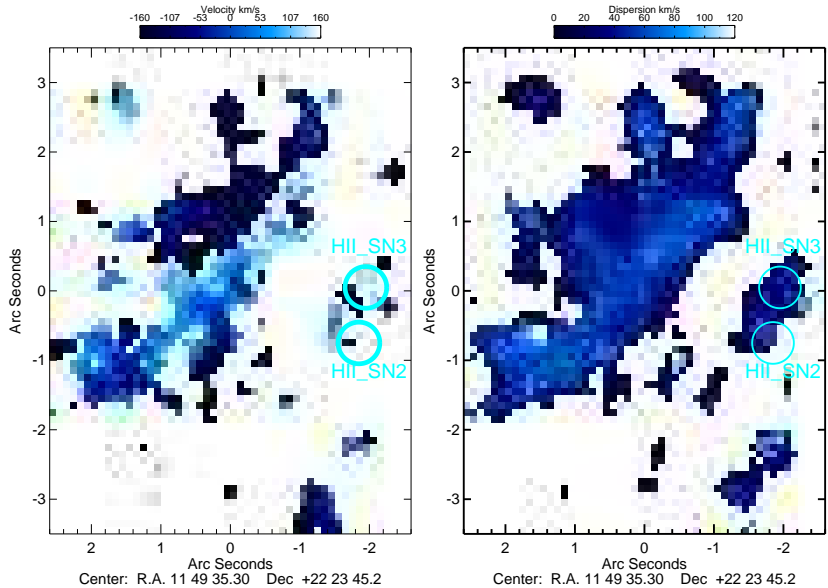}
\caption{Velocity and velocity dispersion maps derived from \ha\ lines from the OSIRS datacube. Two \hii\ regions that are close the SN sites are indicated by cyan circles.   There is no abnormal velocity structure nor  enhanced velocity dispersion  indicative of  strong shocks/winds  at the locations of ``\hii\_{\rm SN3}" and ``\hii\_{\rm SN2}"
}
\label{fig:fig2}
\end{figure*}

\section{METALLICITY}
\subsection{Oxygen Abundance (\oh )}
We use the \nii/\ha\ ratio and the N2 metallicity calibration to calculate oxygen abundance \citep[PP04N2; linear relation of][]{Pettini04}.

In this work, we extract  averaged spectra  for the \hii\ regions (``\hii\_{\rm SN3}" and ``\hii\_{\rm SN2}" on Figure~\ref{fig:fig1}) near the SN site.   
We use the peak of the \ha\ emission as the extraction center.  A radius of 0.$\!\!^{\prime\prime}$3  ($\sim$ 550 pc on  the source plane) is chosen as the size of the \hii\ regions 
 empirically because the S/N of the \ha\ line drops below three outside the radius.   We test different extraction apertures,  shapes, and sizes and find no statistically significant
  differences in the averaged spectra if the extracting aperture is centered   at the peak of \ha\ clumps.  

 The 1$\sigma$ spectra in  Figure~\ref{fig:fig1} are derived from bootstrapping (50000 times; weighted by \ha\ flux) the individual spectrum of each pixel in the extraction aperture.   
  The Gaussian line profile fitting was conducted using a $\chi^2$ minimization procedure with  1/$\sigma$$^{2}$ as the weighting. 
 We then use the statistical error output from the line fitting procedure to estimate the S/N of the flux.  For line fluxes of S/N $<$ 3, we derive the flux upper limit by
 using the 1$\sigma$ spectra and median Gaussian line width (5.27 \AA) of the whole galaxy.  
 The statistical metallicity uncertainties are calculated by propagating the flux errors of the \nii\ and \ha\ lines. The metallicity calibration of the PP04N2 method itself has a 1$\sigma$ dispersion of 0.18 dex \citep{Pettini04}. Therefore, for individual galaxies that have statistical metallicity uncertainties of less than 0.18 dex, 
 we use the systematic errors of 0.18 dex.

In the right panel of Figure~\ref{fig:fig1}, we show the averaged spectra for the \hii\ regions close to  the multiple images of ``SN Refsdal."
Unfortunately, the \nii\  line is below the 3$\sigma$ detection limit for all of the individual \hii\ regions. 
The projected distances from the \hii\ regions to the SN sites are $\sim$ 200 pc and 1.2 kpc for ``\hii\_{\rm SN2} (panel (b))" and ``\hii\_{\rm SN3} (panel (a))" respectively.
 ``\hii\_{\rm SN3}" is less likely to be physically associated with the SN.

To further improve the S/N on the weak \nii\ line, we extract a spectrum averaged across nine \hii\ regions that are at similar galactocentric distances as ``\hii\_{\rm SN2,3}" (the projected distance is $\sim$5-7 kpc on the source-plane; panel (c) in Figure ~\ref{fig:fig1}).  
No \nii\ is detected above an S/N of three in the averaged spectra across the \hii\ regions, consistent with the result from \citet{Yuan11}  that \nii/\ha\ ratios are extremely low at 
the outskirts of the spiral.  For comparison, 
the average spectrum from the innermost annulus of $\sim$ 1.8 kpc (source-plane) is plotted at the  bottom right panel of Figure~\ref{fig:fig1}.

We note that analysis on the source plane gives consistent results though the S/N of the spectra is  slightly downgraded by  1.2 $\times$ due to  further uncertainties in lensing models.  We conclude that  results based on the emission line ratios are not affected by the lensing models. 

The absolute value of oxygen abundance depends on the calibration method \citep{Kewley08}.
The advantages of using PP04N2 for absolute oxygen abundance include the following. 
1)  The \nii/\ha\ ratio is not affected by extinction, and in most cases, is the only line ratio available for high redshift observations. 
2) The absolute value of PP04N2 metallicity is actually closer to direct electron temperature method ({\rm Te}) metallicity than other commonly used strong line diagnostics \citep{Kewley08}.   The  ``{\rm Te}" method metallicity is considered to be a more ``direct" method to calculating the absolute metallicity of galaxies, though it could be  
underestimating  the  value at very high metallicities  \citep[e.g.,][]{Nicholls13}.   
The caveats of using PP04N2 to diagnose the absolute oxygen abundance include the following.
1) The \nii/\ha\ ratio is affected by  ionization parameters and the presence of shocks.   At a fixed metallicity, a higher ionization parameter  produces  lower \nii/\ha\ ratios, and the existence of shock will elevate  \nii/\ha\ ratios \citep[e.g.,][]{Morales13,Rich11,Kewley13a}.
2) The N/O ratio may vary across galaxies, and nitrogen fails to be a good proxy for oxygen abundance when the primary nitrogen production is dominating \citep[e.g.,][]{Morales13}. 
The N2 index should not be used at very low oxygen abundance because the above caveats are most severe at low metallicities   \citep[\oh $\lesssim$ 8.0;][]{Morales13}. 

Keeping the caveats in mind,  the upper limit we derived for the SN site is still useful as a  guide.    The \hii\ region (\hii\_{\rm SN2}) that is $\sim$ 200 pc 
away from the SN has an upper limit metallicity of  \oh $\le$ 8.67. 
Furthermore, the upper limit of  \oh $\le$ 8.11 from the  nine \hii\ region average implies that the oxygen abundance is extremely low: more than $\sim$0.5 dex lower than the central 1.8 kpc and $\sim$ 0.3 dex lower than the average metallicity of the whole galaxy.   
 We can not quantify
the effect of shock or variations in ionization parameters.  However, we do not find enhanced velocity dispersion  indicative of  strong shocks/winds  at the locations of ``\hii\_{\rm SN3}" and ``\hii\_{\rm SN2}" (see Figure~\ref{fig:fig2}).

\begin{figure}[!ht]
\centering
\includegraphics[trim = 2mm 2mm 2mm 3.4mm, clip,width=7.4cm,angle=90]{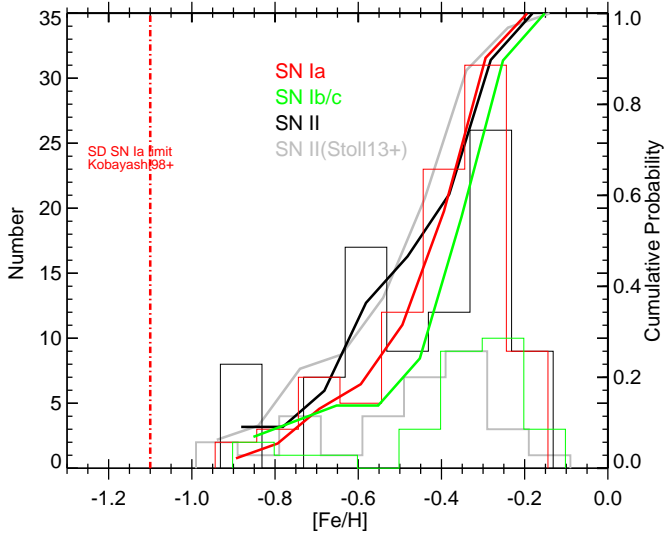}
\caption{Histogram and cumulative distribution of [Fe/H] for the  SNe Ia, II, Ib/c sample from  \citet{Prieto08} and SNe II from \citet{Stoll13}. These local samples show that 
SNe Ib/c occur in higher-metallicity host galaxies than SNe II, whereas there is no statistically significant difference between SNe Ia and SNe II. 
The KS probability tests of two host metallicity samples being drawn from the same distribution are 5\% (II{-}Ib/c), 3\% (Ia{-}Ib/c), and 56\% (II{-}Ia).  The vertical dashed-dotted line 
highlights the theoretical [Fe/H] limit of $\lesssim$-1.1 below which the single-degenerate  path SN Ia is forbidden for M$_{WD}${=}1 M$_{\odot}$ \citet{Kobayashi98}.  
}
\label{fig:fig3}
\end{figure}

\begin{deluxetable*}{lcccc}
\tabletypesize{\scriptsize}
\tablewidth{0pt}
\tablenum{1}
\tablecolumns{5}
\tablecaption{Metallicity and Observed Line Fluxes\label{tab:obs}}
\tablehead{
\colhead{Region} & 
\colhead{Flux \ha}&
\colhead{Flux \nii} &
\colhead{12+log(O/H)PP04N2} &
\colhead{[Fe/H]} \\
\cline{1-5}\\
\colhead{(1)} &
\colhead{(2)} &
\colhead{(3)} &
\colhead{(4)} &
\colhead{(5)} \\
\cline{2-5}\\
}
\startdata
\hii\ SN3(1.2 kpc from SN\tablenotemark{a})&  0.026$\pm$0.004 &$\le$0.004 &$\le$8.45 & $\le$-0.64  \\ 
\hii\ SN2 (200 pc from SN) &     0.010$\pm$0.002     &$\le$0.004&$\le$8.67 &$\le$  -0.36 \\ 
\hii\ SN2,3, \hii\ x1-7 (5-7kpc galactocentric)& 0.022$\pm$0.001 &$\le$0.0009 &$\le$8.11& $\le$-1.06  \\ 
Central 1.8kpc & 0.025$\pm$0.002  &0.011$\pm$0.0012 &8.69$\pm$0.05\tablenotemark{b}&-0.34  \\ 
Whole galaxy\tablenotemark{c}& 0.024$\pm$0.0008  &0.004$\pm$0.001 &8.43$\pm$0.05\tablenotemark{b}&-0.66   \\ 
\enddata
\tablecomments{
(1) Region where the spectra are averaged, labeled in Figure 1.
(2), (3)  Averaged observed flux density over respective spatial regions; in units of 10$^{-16}$ ergs~s$^{-1}$~cm$^{-2}$~per spaxel (0.$\!\!^{\prime\prime}$1$^{2}$). 1$\sigma$ upper limit is derived for line detection below an S/N of 5.
(4) Metallicity based on the PP04N2 method.
(5) [Fe/H] inferred from oxygen abundance (see Section 3.2)
\tablenotemark{a} Intrinsic distance corrected for lensing, uncorrected for orientation. Typical uncertainty of the distance from lens model is $\sim$ 20\%.
\tablenotemark{b} This is the statistical error.  Systematic error from the N2-dex method is $\sim$ 0.18 dex.
\tablenotemark{c} The average flux density of the whole galaxy is derived using the same method as the \hii\ regions, with an aperture size that includes the whole galaxy. 
 The total SFR of the  galaxy  after lensing and dust correction is $\sim$ 1.2 M$_{\odot}$ yr$^{-1}$  \citep{Livermore15}. 
}
\end{deluxetable*}

\subsection{ Iron Abundance [Fe/H]}
Because it is iron rather than oxygen that provides the dominant opacity for the white dwarf winds in the late stage of stellar evolution \citep{Kobayashi98}, we 
 infer stellar iron abundance ([Fe/H]) using gas-phase oxygen abundance (\oh).     
  
The [O/Fe] ratios have a well-known relation with [Fe/H] in the solar neighborhood, where [O/Fe] is constant at [Fe/H] $\lesssim -1$ and linearly decreases at [Fe/H] $\gtrsim -1$ 
 \citep[e.g.,][]{Cayrel04}. The plateau [Fe/H]  is determined from the IMF weighted yields of CC SNe (SNII and Ibc) and is 0.45,   consistent with observations of stars with 3D/NLTE corrections  (Kobayashi et al. 2006). With this constant, the upper limit of [Fe/H] of the near the site of "SN Refsdal"  is $-1.06$ for  \oh $\le$ 8.11 (nine \hii\ average).

We use the prescriptions in \citet{Stoll13} for the conversion:  $\rm{[Fe/H]} = -11.2 + 1.25(12+\log(O/H))_{PP04N2}$.
We list the converted [Fe/H] values for SN sites in Table 1. Here, we adopt the solar metallicity \ohe 8.69 \citep{Asplund09}. 
If  we use other values of solar metallicity, e.g., \ohe 8.86 from \citet{Delahaye06},  the  [Fe/H]  ratio is  $\sim$ 0.2 dex lower.
This conversion is based on fitting to the  linear correlation between [Fe/H] and [O/H], which is valid above [Fe/H] {=} $-1$.

Note that oxygen is one of the $\alpha$ elements (O, Mg, Si, S, and Ca), and observed [$\alpha$/Fe] ratios do depend on the system; [$\alpha$/Fe] is higher at the given [Fe/H] in the Galactic bulge and thick disk, while [$\alpha$/Fe] is lower at the given [Fe/H] in Magellanic Clouds and dwarf spheroidal galaxies \citep[e.g,][]{Kobayashi06}. Damped Ly$\alpha$ systems, which may correspond to the outskirts of spiral galaxies at high redshifts, show low [$\alpha$/Fe] ratios, although this depends on the dust depletion correction 
\citep[e.g.,][]{Wolfe05}. This [O/Fe] variation is not taken into account in the column 5 of Table 1.  It is unknown if the relation between O and Fe stellar abundances can be used to convert the gas abundance.   We therefore caution the limitation of this method.

\section{DISCUSSION}
Current observational studies on local galaxies have shown some patterns between the types of SN progenitor and the host galaxy metallicities.
For example, by directly comparing the central oxygen abundance of the hosts of SNe Ib/c and SNe Ia with SNe II in SDSS galaxies, 
 \citet{Prieto08} found strong evidence that SNe Ib/c occur in higher-metallicity host galaxies than SNe II, whereas there is no statistically significant
 difference between SNe Ia and SNe II.  Similar conclusions are also found in other studies that have metallicities directly measured at the SN sites  \citep{Modjaz08,Stoll13}. 
 Although normal SN Ia have not been found at [Fe/H] $\lesssim -1$; the metallicity of one of the most metal-poor SNIa host galaxies is 12 + log(O/H){=}8.01 \citep[with the R23 method;][]{Childress11}.

 To illustrate this, in Figure~\ref{fig:fig3} we plot the histogram and cumulative distribution of [Fe/H] for the SNe Ia, II, Ib/c sample from  \citet{Prieto08}. We derive the oxygen abundance 
 using PP04N2 and convert it to [Fe/H] using the same methods described in Section 3.2.   We also plot the sample of \citet{Stoll13} for Type II SN progenitor regions from a 
 single-source survey that minimized the host galaxy selection bias.    The metallicity distributions of SNe II from the two samples agree well, and there is no
 distinguishable difference between Types Ia and II.  Type Ib/c show a preference toward higher [Fe/H]. 
 
 The vertical dashed-dotted line in Figure~\ref{fig:fig3} marks the theoretical [Fe/H] limit of $\sim$$-1.1$, below which the single-degenerate 
 path SN Ia is forbidden \citep{Kobayashi98}.  Observations have shown evidence that such  a boundary for SNe Ia exists. 
For example, \citet{Rodney14}  found a low rate of SNe Ia at $z >$ 1 which could be explained by a population of single-degenerate-dominated Type Ia at high redshift. 
Note that certain types of SNe may occur only for low-metallicity progenitor stars \citep[e.g.,][]{Langer07}. \citet{Childress11} found super-Chandrasekhar candidates in metal-poor environments.

The inferred [Fe/H] upper limit of $\lesssim -1$ using the nine \hii\ region average is close to the boundaries of the \citet{Kobayashi98} limit, 
causing mild tension for ``SN Refsdal"  to be a  normal Type Ia. 
The [Fe/H] of ``SN Refsdal"  is at the low end tail of the [Fe/H] distribution of local SNe. Given that Type Ib/c usually favor high-metallicity environments, it could imply that
``SN Refsdal" is less likely to be a Ib/c in a relative sense.   However, the SN type of  ``SN Refsdal"  is still unknown.  Future follow-up observations will determine 
the SN type of `SN Refsdal"  and thus lead to more meaningful discussion on the metallicity constraints.

\section{CONCLUSION}
We use our updated near-infrared IFS data to place constraint on the local metallicity  near the site of ``SN Refsdal." 
 Based on our  updated near-infrared IFS data,  the gas-phase metallicity averaged in an intrinsic radius of $\sim$ 550 pc surrounding  an \hii\ region  $\sim$ 200 pc away from the SN site is   12 + log(O/H)$_{\rm PP04N2}$ $\le$ 8.67.  The  metallicity averaged over nine \hii\ regions at    similar galactocentric distances  ($\sim$5-7 kpc) as ``SN Refsdal"  is constrained to be 12 + log(O/H)$_{\rm PP04N2}$ $\le$ 8.11.    
  
Given the fortuitous discovery of ``SN Refsdal" in an advantageously lensed face-on spiral, this is by far the best observational constraint we can set on the local metallicity environment of an SN site at redshift $z>1$.   The next generation of 30 m telescopes, such as TMT and GMT, are ideal for pursuing the \hii\ region metallicity of the host galaxy of `SN Refsdal."  Future spectroscopic and imaging  follow-up work  is required to conclude the supernova type of  ``SN Refsdal."   Our metallicity constraints provide an important  parameter to understand the  physical origin of  ``SN Refsdal."

\acknowledgments 
We thank the  referee for insightful criticism  that has significantly improved this paper.
T.Y. thanks the useful discussion with Mike Childress, Fang Yuan, Richard Scalzo, and  D.-C. Nicholls.  L.K. acknowledges the support of an ARC Future Fellowship and ARC Discovery Project DP130103925.   The data presented in this paper were obtained at the
W.M. Keck Observatory. The authors wish to recognize and
acknowledge the very significant cultural role and reverence that the
summit of Mauna Kea has always had within the indigenous Hawaiian
community. 


\end{document}